\documentclass[twocolumn,amsmath,amssymb,aps,pra,]{revtex4-1}
\usepackage{dcolumn}
\usepackage{bm}
\usepackage[colorlinks=true,urlcolor=blue,linkcolor=blue,citecolor=blue]{hyperref}
\usepackage{color}
\usepackage{graphics}
\usepackage{graphicx}
\usepackage{hyperref}

\begin{document}

\title{Dynamics of an unbalanced two-ion crystal in a Penning trap for application in optical mass spectrometry}

\author{M.J.~Guti\'errez$^{1}$}\email[]{mjgutierrez@ugr.es.}\thanks{This work is part of the PhD thesis of M.J. Guti\'errez.}
\author{J.~Berrocal$^{1}$} 
\author{F.~Dom\'inguez$^1$}
\author{I.~Arrazola$^{2}$}
\author{M.~Block$^{3,4,5}$}
\author{E.~Solano$^{2,6,7}$}
\author{D.~Rodr\'iguez$^{1,8}$}\email[]{danielrodriguez@ugr.es}
\affiliation{
$^1$Departamento de F\'isica At\'omica, Molecular y Nuclear, Universidad de Granada, 18071 Granada, Spain \\
$^2$Department of Physical Chemistry, University of the Basque Country UPV/EHU, Apartado 644, E-48080, Bilbao, Spain\\
$^3$Institut für Kernchemie, Johannes Gutenberg-Universität Mainz, D-55099, Mainz, Germany \\
$^4$GSI Helmholtzzentrum für Schwerionenforschung GmbH, D-64291, Darmstadt, Germany \\
$^5$Helmholtz-Institut Mainz, D-55099, Mainz, Germany \\
$^6$IKERBASQUE, Basque Foundation for Science, Maria Diaz de Haro 3, E-48013, Bilbao, Spain\\
$^7$Department of Physics, Shanghai University, 200444 Shanghai, People's Republic of China\\
$^8$Centro de Investigaci\'on en Tecnolog\'ias de la Informaci\'on y las Comunicaciones, Universidad de Granada, 18071 Granada, Spain
}

\date{\today}

\begin{abstract}
In this article, the dynamics of an unbalanced two-ion crystal comprising the 'target' and the 'sensor' ions confined in a Penning trap has been studied. First, the low amplitude regime is addressed. In this regime, the overall potential including the Coulomb repulsion between the ions can be considered harmonic and the axial, magnetron and reduced-cyclotron modes split up into the so-called 'stretch' and 'common' modes, that are generalizations of the well-known 'breathing' and 'center-of-mass' motions of a balanced crystal made of two ions. By measuring the frequency modes of the crystal and the sensor ion eigenfrequencies using optical detection, it will be possible to determine the target ion's free-cyclotron frequency. The measurement scheme is described and the non-harmonicity of the Coulomb interaction is discussed since this might cause large systematic effects. 
\end{abstract}

\maketitle


\section{\label{sec::intro} Introduction}
A precise determination of atomic and nuclear masses of exotic particles is of fundamental interest in many areas, for example in nuclear physics and in neutrino physics \cite{Bloc2010,Elis2015}. The determination of nuclear binding energies of exotic nuclides has contributed to a better understanding of their nuclear structure by identifying the changes in shell structure or by finding the onset of deformation (see e.g. \cite{Mina2012}). Superheavy elements (SHE) are among of the most challenging objects for such investigations since they can only be produced in quantities of few atoms at a time in only four facilities worldwide, and they are often short-lived \cite{Munz2015,Munz2015b,Mori2015,Ogan2015}. Among these large-scale facilities, there is just one Penning-trap system, at GSI-Darmstadt \cite{Bloc2005}, which allows direct mass measurements by means of Penning traps \cite{Bloc2019}. However, the minute production yields for SHE ($Z\geq 104$) call for measurement techniques of utmost sensitivity that are universal and fast. Ideally, precise mass values should be obtained from a single ion for ions of a wide mass range. In addition, in case of a non-destructive measurement technique, additional observables such as nuclear decay modes and half-lives can be obtained from the very same ion. Such an approach would also benefit precise mass measurements that are required in the context of neutrino physics where often the mass of rare isotopes, available only in limited samples, are demanded. The limitation in sensitivity of the Phase-Imaging Ion-Cyclotron-Resonance (PI-ICR) technique, currently in use at GSI-Darmstadt, to about 10 ions \cite{Elis2013} motivated the development of new techniques \cite{Rodr2012,Lohs2019}. 

In this publication, we present the results from numerical and analytical calculations of the dynamics of an unbalanced two-ion crystal in a 7-tesla Penning trap \cite{Guti2019}. The outcomes from the calculations can be combined with the optical response of the sensor ion after applying external fields in resonance with the motional modes of the crystal. The latter has already been studied using a Paul trap with rotational symmetry \cite{Domi2017,Domi2017b}. The proposed method is universal (with respect to the mass-to-charge ratio), it can be used in broad band mode, and offers a permanent monitoring of the motion of the sensor ion in the trap, allowing for a precise quantification of possible systematic effects arising from trap-electrodes imperfections or misalignments with respect to the magnetic field. Besides sensitivity, using the optical control, an improvement in precision and accuracy is envisaged, so as to carry out mass measurements, for example on $^{187}$Re$^+$ and $^{187}$Os$^+$ \cite{Nest2014}, with a competitive level of accuracy in order to contribute to the determination of the mass of the electron antineutrino. Such ions have been already produced in the TRAPSENSOR facility \cite{Corn2013,Guti2019}. Extending this measurement method to other ion pairs like $^{163}$Ho$^+$~-$^{163}$Dy$^+$ \cite{Elis2015} will be also possible.

\section{\label{sec::dynamics}Penning trap dynamics}

\begin{figure}
\centering
\includegraphics[width=0.5\textwidth]{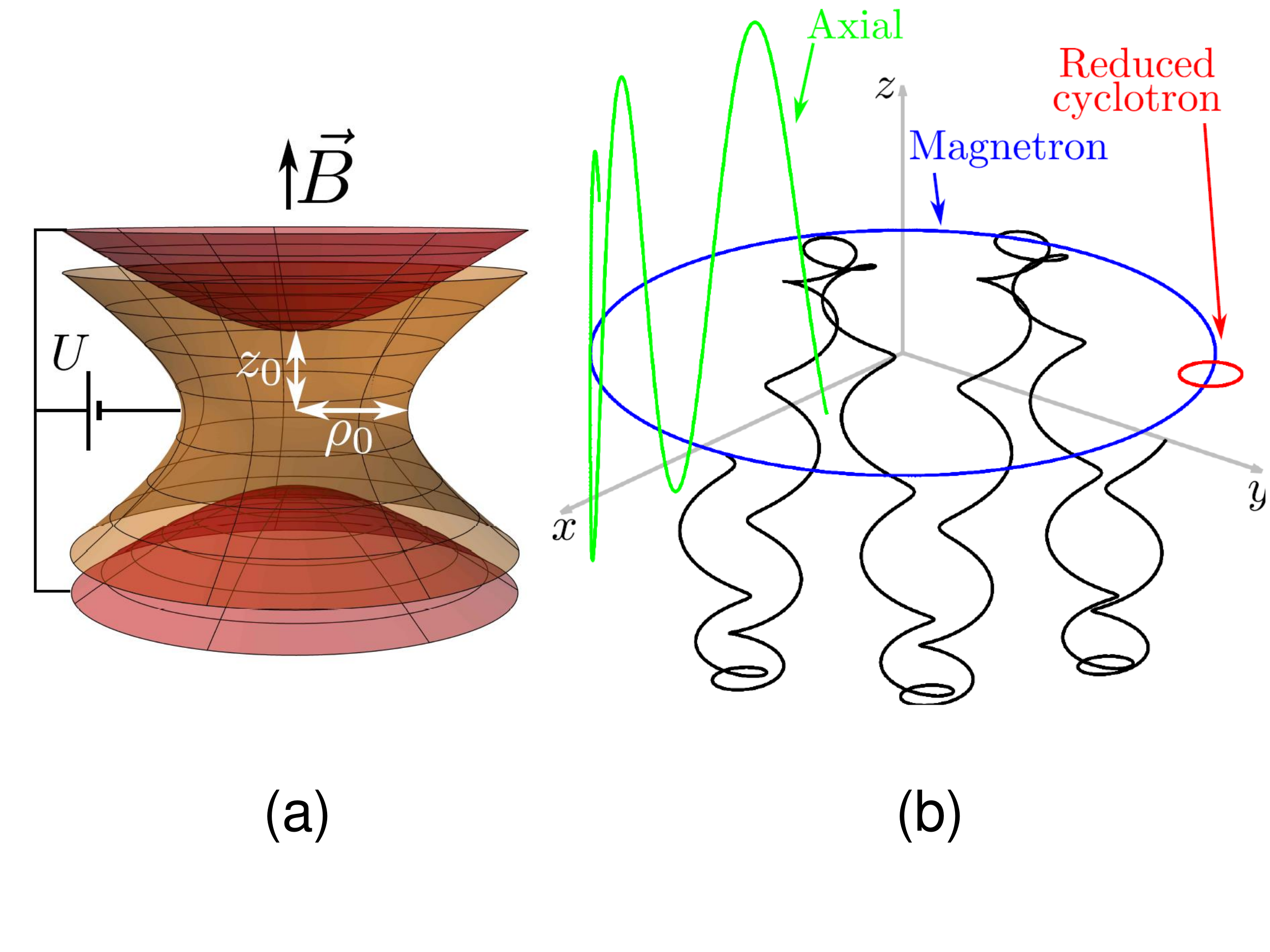}
\vspace{-1.4cm}
\caption{Left: original shape of a Penning trap (revolution hyperboloid). The yellow electrode is referred to as \emph{ring}, and the red ones are known as \emph{endcaps}. The so-called \emph{characteristic distance} of the trap is given by $d_0^2=\left(z_0^2+\rho_0^2/2\right)/2$. Right: Normal modes of a single ion stored  in a Penning trap. The black solid-line depicts the overall motion.}
\label{fig::penningtrap}
\end{figure}

A Penning trap (see the left panel of Fig.~\ref{fig::penningtrap}), uses the combined effect of a strong homogeneous magnetic field, $ \vec B = B \hat z $, and a quadrupolar electrostatic field $ V = U/4d_0^2 \cdot \left( 2z^2 - x^2 - y^2 \right)$, to confine a charged particle or a crystal \cite{Brow1986}. In this section the dynamics of the unbalanced two-ion crystal in a Penning trap is presented, after introducing the dynamics of the single trapped ion and obtaining the general equations in the low-amplitude regime. Solving these equations numerically for a particular case allows characterizing the frequency shifts arising from the non-harmonic nature of the Coulomb repulsion. Previous work on ion crystals formed by identical ions stored in the same Penning trap, has been carried out at Imperial College \cite{Thom1997,Mava2014}, while ultra-accurate mass spectrometry using two different ions stored simultaneously in a large magnetron orbit, to minimize their electrostatic interaction, was developed at Massachusetts Institute of Technology \cite{Corn1992,Rain2004}. This Penning trap was relocated afterwards to Florida State University \cite{Moun2009}. 

The motion of a single ion confined in an ideal Penning trap is well known \cite{Brow1986}. The only force with a non-zero projection along the revolution axis is that of the electrostatic field. Since the field is quadrupolar, the resulting motion is harmonic with an oscillation frequency
\begin{equation}
\omega_z = 2\pi \nu_z = \sqrt{ \frac{qU}{md_0^2} } \,.
\end{equation}
The motion in the radial plane is an epicyclic orbit with frequencies
\begin{equation}
\nu_{c'/m}= \frac{\nu_c}{2} \left[ 1 \pm \sqrt{ 1 - 2 \left( \frac{\nu_z}{\nu_c} \right)^2 } \right] = \frac{\nu_c}{2} \pm \omega_1 \,.
\label{eq::radialfreqs}
\end{equation}
The subscript $c'$ (corresponding to the $+$ sign) is associated to the so-called reduced-cyclotron motion, whose frequency is very close to the free cyclotron frequency of the ion in the magnetic field, $\omega_c = 2\pi \nu_c = qB/m$. The subscript $m$, on the other hand, corresponds to the magnetron motion, whose frequency is much lower than all the other involved frequencies under normal operation conditions.

The relationship between $\nu_c$ and the ion's mass depends only on the magnetic field, thus the precise determination of $\nu_c$ is the most common way to perform high-precision mass measurements. Since the ions do not oscillate with this frequency, relationships between the motional frequencies and the true cyclotron frequency are utilized, namely
\begin{equation}
\nu_c = \nu_{c'} + \nu_m
\end{equation}
and
\begin{equation}
\nu_c^2 = \nu_{c'}^2 + \nu_z^2 + \nu_m^2 \,.
\label{eq::invariance}
\end{equation}
These can be easily derived from Eq.~(\ref{eq::radialfreqs}). The latter involves the measurement of an additional eigenfrequency, $\nu_z$. However, it provides an advantage: Eq.~(\ref{eq::invariance}), known as \emph{invariance theorem}, holds for \emph{real} Penning traps \cite{Brow1982}, i.e., traps with a small misalignment between electric and magnetic field or with electric field imperfections.



In a crystal made of two different ions, one defines the mass ratio $\mu=m_t/m_s$, where the subscript $t$ denotes the 'target' ion or 'ion of interest' and $s$ the 'sensor' ion. The axial and true cyclotron frequencies of the target ion are $\omega_{zt} = \omega_0/\sqrt{\mu} $ and $ \omega_{ct} = \omega_{cs} / \mu $, respectively, with $\omega_0$ and $\omega_{cs}$ being the axial and true cyclotron frequency of the sensor ion. The orientation of the crystal can be axial or radial. This depends on frequency ratios. For example, if the crystal comprises two identical ions with masses $m_t$ (balanced crystal), the axial orientation is energetically favorable if \cite{Thom1997}
\begin{equation}
m_t < \frac{1}{6}\frac{qB^2d_0^2}{U},
\end{equation}
where the right part of the equation depends only on the trap parameters.

\begin{figure}
\centering
\includegraphics[width=0.5\textwidth]{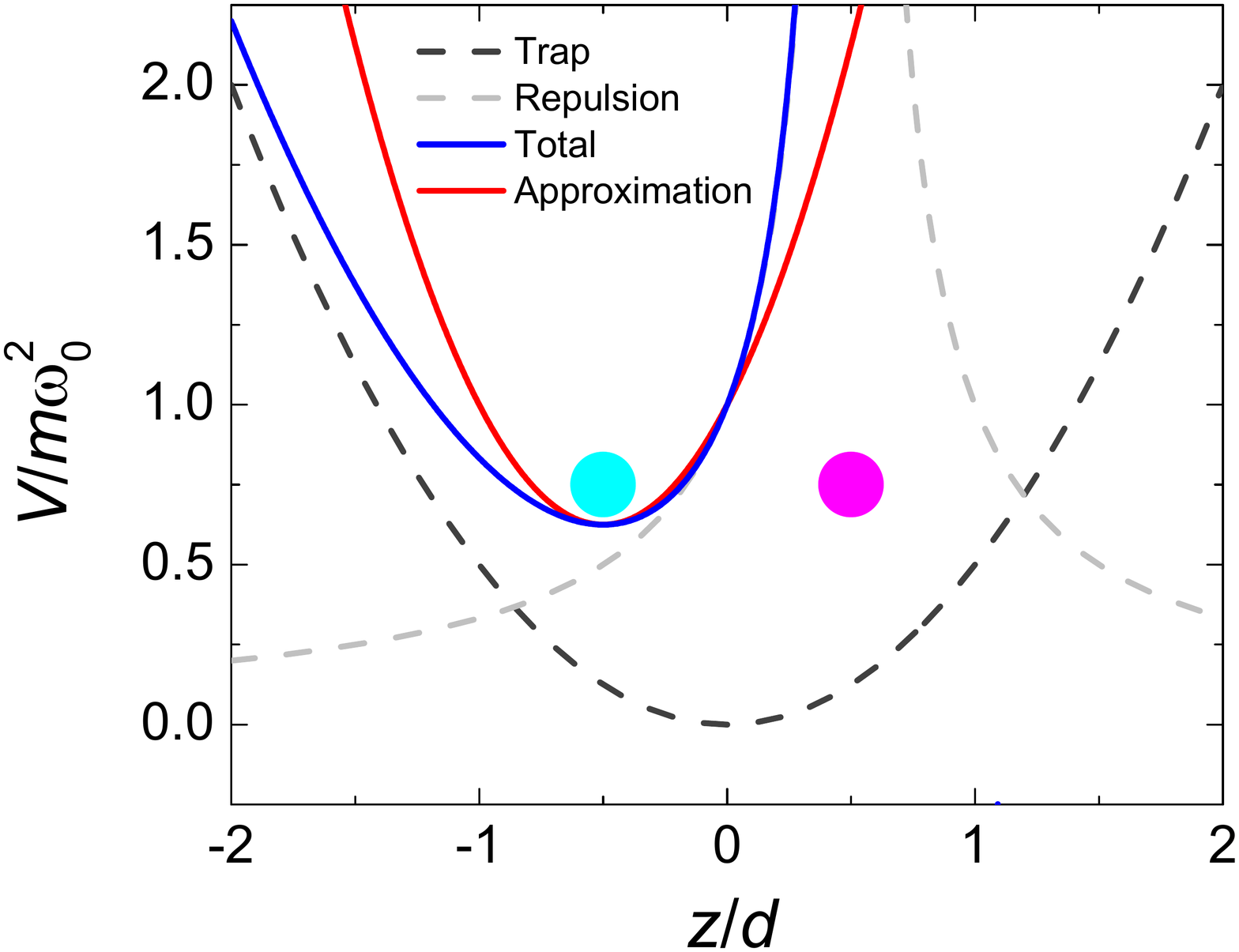}
\vspace{-0.6cm}
\caption{Electric potentials as seen by the ion at $z=-d/2$ (colored in cyan) along the trap axis. The repulsion term is due to the presence of a second (pink-colored) ion.}
\label{fig::axial1}
\end{figure}

In order to solve the equations for the unbalanced crystal (Appendix~\ref{Equations_of_motion}), we will work in the small amplitude regime, i.e., when the displacement from the equilibrium position is $<<d$ (see Fig.~\ref{fig::axial1}). Under these conditions, the Coulomb interaction can be considered harmonic. In this case the Coulomb force is Taylor expanded to first order and the resulting equations are
\begin{eqnarray}
\begin{cases}
\displaystyle{\ddot{x}_t = \frac{1}{\mu} \omega_0^2 x_t + \frac{1}{\mu} \omega_{cs} \dot{y}_t - \frac{1}{2} \frac{1}{\mu} \omega_0^2 x_s}
\\
\displaystyle{\ddot{y}_t = \frac{1}{\mu} \omega_0^2 y_t - \frac{1}{\mu} \omega_{cs} \dot{x}_t - \frac{1}{2} \frac{1}{\mu} \omega_0^2 y_s}
\\
\displaystyle{\ddot{z}_t = -2 \frac{1}{\mu} \omega_0^2 z_t + \frac{1}{\mu} \omega_0^2 z_s}
\\
\displaystyle{\ddot{x}_s = \omega_0^2 x_s + \omega_{cs} \dot{y}_s - \frac{1}{2} \omega_0^2 x_t}
\\
\displaystyle{\ddot{y}_s = \omega_0^2 y_t - \omega_{cs} \dot{x}_s - \frac{1}{2} \omega_0^2 y_t}
\\
\displaystyle{\ddot{z}_s = -2 \omega_0^2 z_s + \omega_0^2 z_t}
\end{cases} \,.
\label{eq::linearized}
\end{eqnarray}

In this approximation, the coupling between radial and axial motions disappears, and thus the equations in the radial plane and axial direction can be solved separately.


\subsection{Axial motion}

\begin{figure}
\centering
\includegraphics[width=0.5\textwidth]{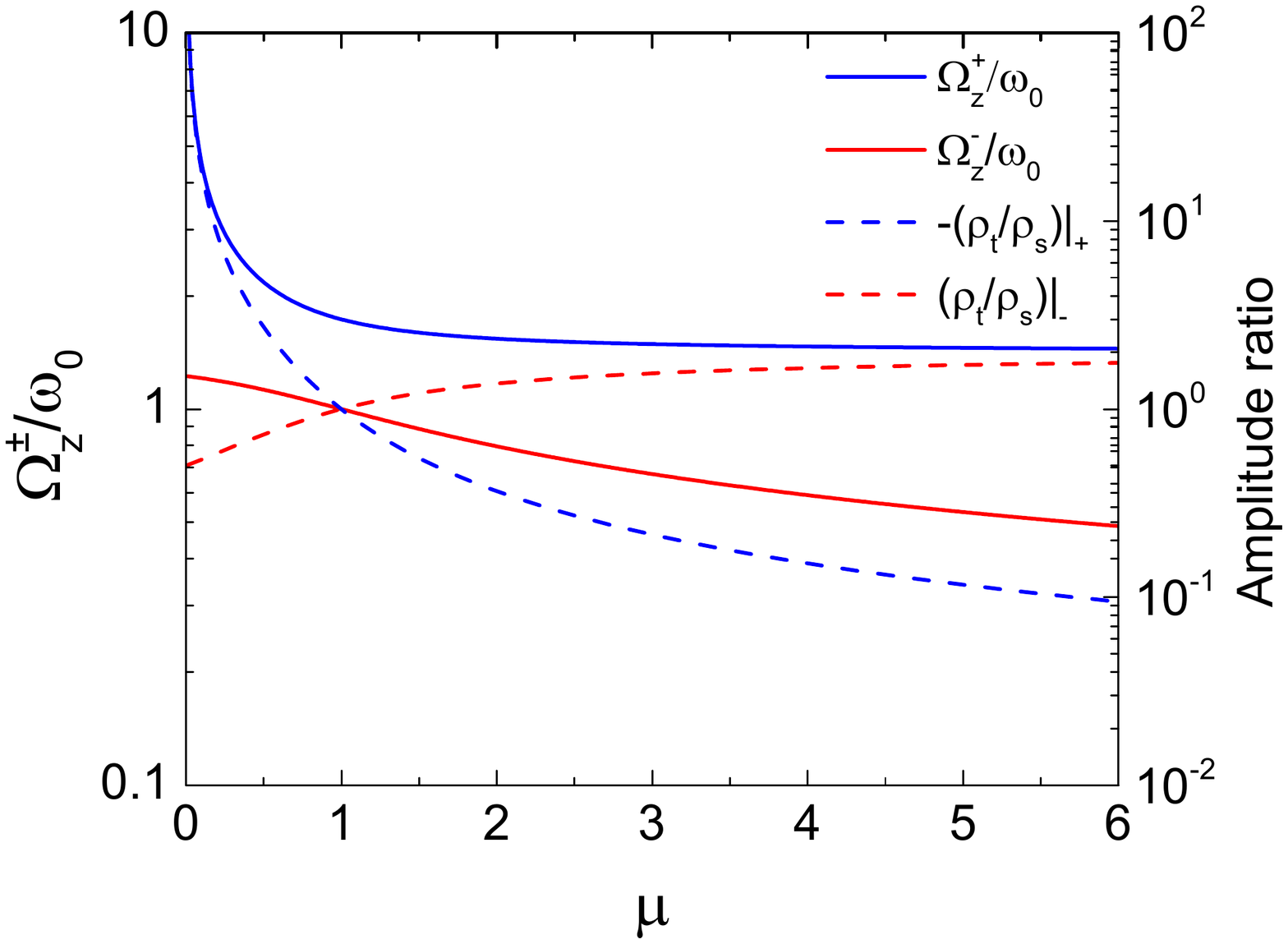}
\vspace{-6.5cm}
\caption{Eigenvalues and eigenvectors for the two-ion crystal in the axial degree-of-freedom. The solid lines correspond to the axial frequencies in units of the axial frequency of the sensor ion. Note that the \emph{stretch} frequency is almost constant for large values of $\mu$; the same happens for the \emph{common} mode when $\mu$ is small. The dashed lines correspond to the ions' amplitude ratio (target over sensor).}
\label{fig::axial}
\end{figure}

The equations describing the axial motion can be written in matrix form as
\begin{equation}\label{matrix-axial}
\left(
\begin{matrix}
\ddot{z}_t \\
\ddot{z}_s
\end{matrix}
\right)
=
\omega_0^2 \left(
\begin{matrix}
\frac{-2}{\mu} & \frac{1}{\mu} \\
1 & -2
\end{matrix}
\right)
\cdot
\left(
\begin{matrix}
z_t \\
z_s
\end{matrix}
\right) \,.
\end{equation}
The normal modes and frequencies are obtained by diagonalizing the coefficient matrix, yielding
\begin{equation}
\ddot{Z}_\pm = - \left( \Omega^\pm_z \right)^2 Z_\pm \,,\,\,
\left( \Omega^\pm_z \right)^2 = \omega_0^2 \left[ 1 + \frac{1}{\mu} \pm \sqrt{ 1 + \frac{1}{\mu^2} - \frac{1}{\mu} } \right] \,.
\label{eq::axial-eigenmodes}
\end{equation}
Here, $+$ and $-$ stand for \emph{stretch} and \emph{common} mode, respectively. They are related to the amplitudes of the ions by
\begin{equation}
Z_\pm = \left( 1 - \frac{1}{\mu} \mp \sqrt{ 1 + \frac{1}{\mu^2} - \frac{1}{\mu} } \right) z_t + z_s \,,
\end{equation}
where a global normalization factor has been omitted. Thus, the ions move in phase for the common mode, and out of phase for the stretch mode. The lighter ion moves with more amplitude than the heavier one in this mode, and with more amplitude in the common one as shown in Fig.~\ref{fig::axial}. These results naturally agree with those obtained using a Paul trap \cite{Mori2001}.

\subsection{Radial motion}

The radial part of Eq.~(\ref{eq::linearized}) can be written as 

\begin{multline}
\left(
\begin{matrix}
\ddot{u}_t \\
\ddot{u}_s
\end{matrix}
\right)
=
-i \omega_{cs}
\left(
\begin{matrix}
\frac{1}{\mu} &  0  \\
       0      &  1
\end{matrix}
\right)
\cdot
\left(
\begin{matrix}
\dot{u}_t \\
\dot{u}_s
\end{matrix}
\right)
+
\omega_0^2
\left(
\begin{matrix}
\frac{1}{\mu} & -\frac{1}{2}\frac{1}{\mu} \\
-\frac{1}{2}  &            1 
\end{matrix}
\right)
\cdot
\left(
\begin{matrix}
u_t \\
u_s
\end{matrix}
\right)
= \\
= M_{dot}
\cdot
\left(
\begin{matrix}
\dot{u}_t \\
\dot{u}_s
\end{matrix}
\right)
+
M
\cdot
\left(
\begin{matrix}
u_t \\
u_s
\end{matrix}
\right)
\,.
\label{eq::radial-matrix}
\end{multline}
\noindent after the usual change of variable $ u_{t,s} = x_{t,s} + i y_{t,s} $. The easiest way to find the normal modes and frequencies of the system is to find a base where $M$ and $M_{dot}$ are simultaneously diagonal, thus, both matrices have to commute. However,
\begin{multline}
\frac{i}{ \omega_{cs}\omega_0^2 } \left[ M_{dot}, M \right] = \\
=
\frac{1}{2}\frac{1}{\mu} \left( \mu-1 \right)
\left(
\begin{matrix}
 0 & -\frac{1}{\mu} \\
 1 &       0        \\
\end{matrix}
\right)
= 0 \Leftrightarrow \mu = 1 \,.
\label{eq::commutation}
\end{multline}

\noindent This implies that the radial motion can only be studied using this procedure when $\mu=1$ (equal masses). Although this case is of no interest for mass spectrometry, its study gives insight on the motion of two simultaneously trapped ions. Diagonalizing $M$ in Eq.~(\ref{eq::radial-matrix}) yields the eigenvalues
\begin{equation}
\Lambda_\pm = \omega_0^2 \frac{2 \pm 1}{2} \label{aux}
\end{equation}
and eigenvectors
\begin{equation}
U_\pm = u_t \mp u_s \,.
\end{equation}

\noindent $U_-$ corresponds to a \emph{center-of-mass} motion in the radial plane ($U_+ = 0 \Rightarrow u_t = u_s$). $U_+$ on the other hand, is a \emph{breathing} motion ($U_- = 0 \Rightarrow u_t = - u_s $). Equation~(\ref{eq::radial-matrix}) can be written, in its diagonalized form, as

\begin{equation}
\begin{cases}
\ddot U_+ = - i \omega_{cs} \dot U_+ + \frac{3}{2} \omega_0^2 U_+
= - i \omega_{cs} \dot U_+ + \Lambda_+ U_+  \\
\ddot U_- = - i \omega_{cs} \dot U_- + \frac{1}{2} \omega_0^2 U_-
= - i \omega_{cs} \dot U_- + \Lambda_- U_-  \\
\end{cases}
\end{equation} 

\begin{figure}
\centering
\includegraphics[width=0.5\textwidth]{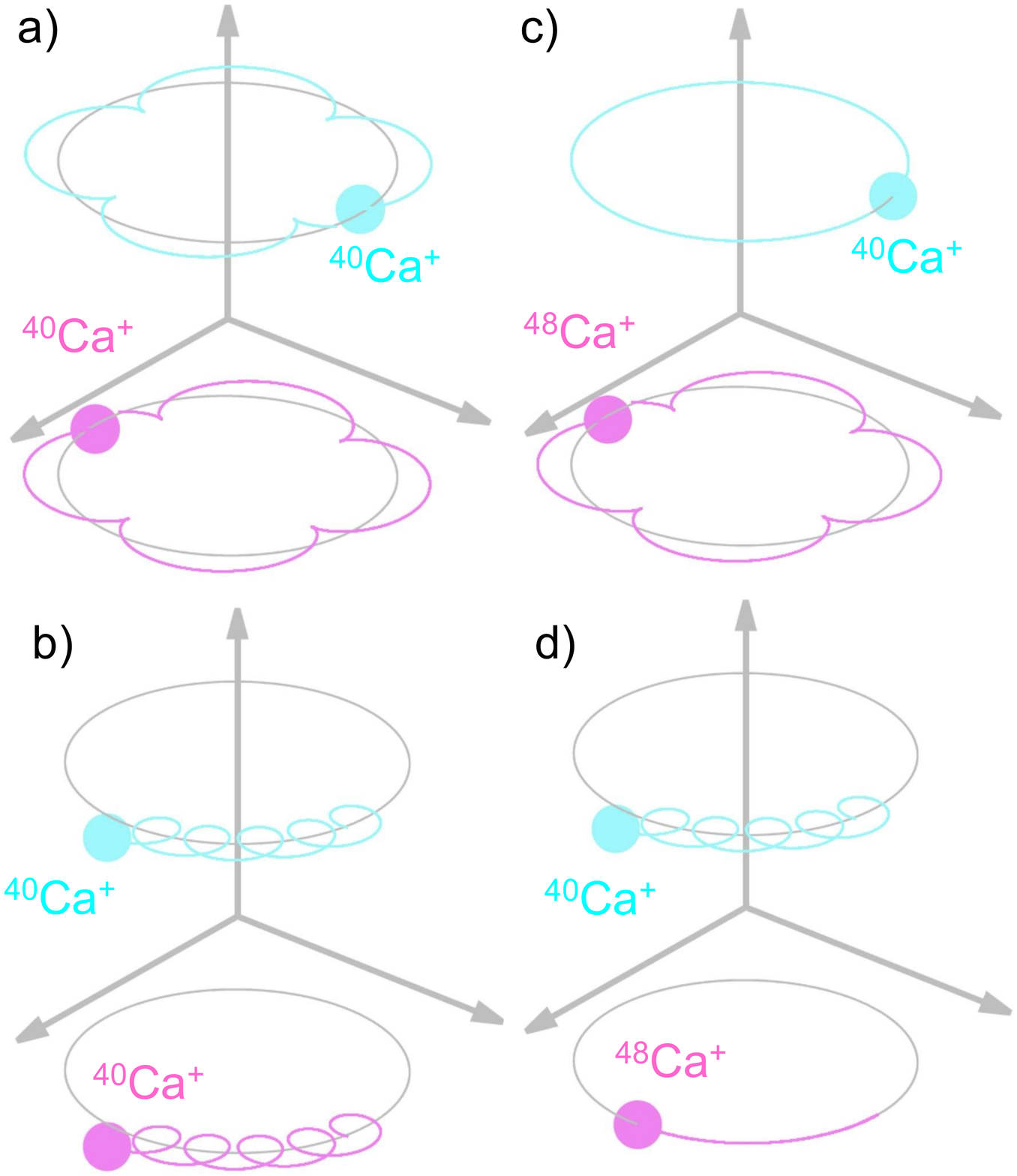}
\vspace{-2.5cm}
\caption{a) Balanced crystal moving in the breathing (magnetron and reduced-cyclotron) modes. b) Balanced crystal moving in the center-of-mass modes. c) Unbalanced crystal moving in the stretch mode (magnetron and reduced-cyclotron). d) Unbalanced crystal moving in the center of mass mode.}
\label{fig::homogeneous-radial}
\end{figure}

\noindent Each of these is equivalent to the equation of motion of a single ion, and can be solved to obtain the motional frequencies. These are, for the $+$ mode,
\begin{equation}
\Omega^+_{c',m} =
\frac{ \omega_{cs} }{2} \left[ 1 \pm \sqrt{ 1 - 2 \left( \frac{\sqrt{3/2} \cdot \omega_0}{\omega_{cs}} \right)^2 } \right] \,.
\end{equation}
Again, the subscript $c'$ corresponds to the $+$ sign in front of the square root, and the subscript $m$ to the minus sign. Similarly, for the $-$ mode,
\begin{equation}
\Omega^-_{c',m} =
\frac{ \omega_{cs} }{2} \left[ 1 \pm \sqrt{ 1 - 2 \left( \frac{\sqrt{1/2} \cdot \omega_0}{\omega_{cs}} \right)^2 } \right] \,,
\end{equation}
which can be rewritten as 
\begin{equation}
\Omega^\pm_{c',m} = \frac{ \omega_{cs} }{2} \left[ 1 \pm \sqrt{ 1 - 2 \left( \frac{\Omega_z^\pm}{\omega_{cs}} \right)^2 } \right] \,.
\label{eq::radial-freqs-homogeneous}
\end{equation}
where $\Omega_z^\pm = \sqrt{ \frac{2 \pm 1 }{2} }\omega_0 $ for two identical ions (Eq.~(\ref{eq::axial-eigenmodes})). This equation is very similar to Eq.~(\ref{eq::radialfreqs}) and yields the radial frequencies of a balanced crystal as a function of the single-ion cyclotron frequency and the axial frequencies of the crystal. Figure~\ref{fig::homogeneous-radial}(a)~and~(b) depicts the radial motion of this crystal. The higher frequency for the reduced-cyclotron motion corresponds to the center-of-mass motion. This is different compared to the axial and magnetron motions, where the higher frequencies correspond to the breathing mode.

\begin{figure}
\centering
\includegraphics[width=0.5\textwidth]{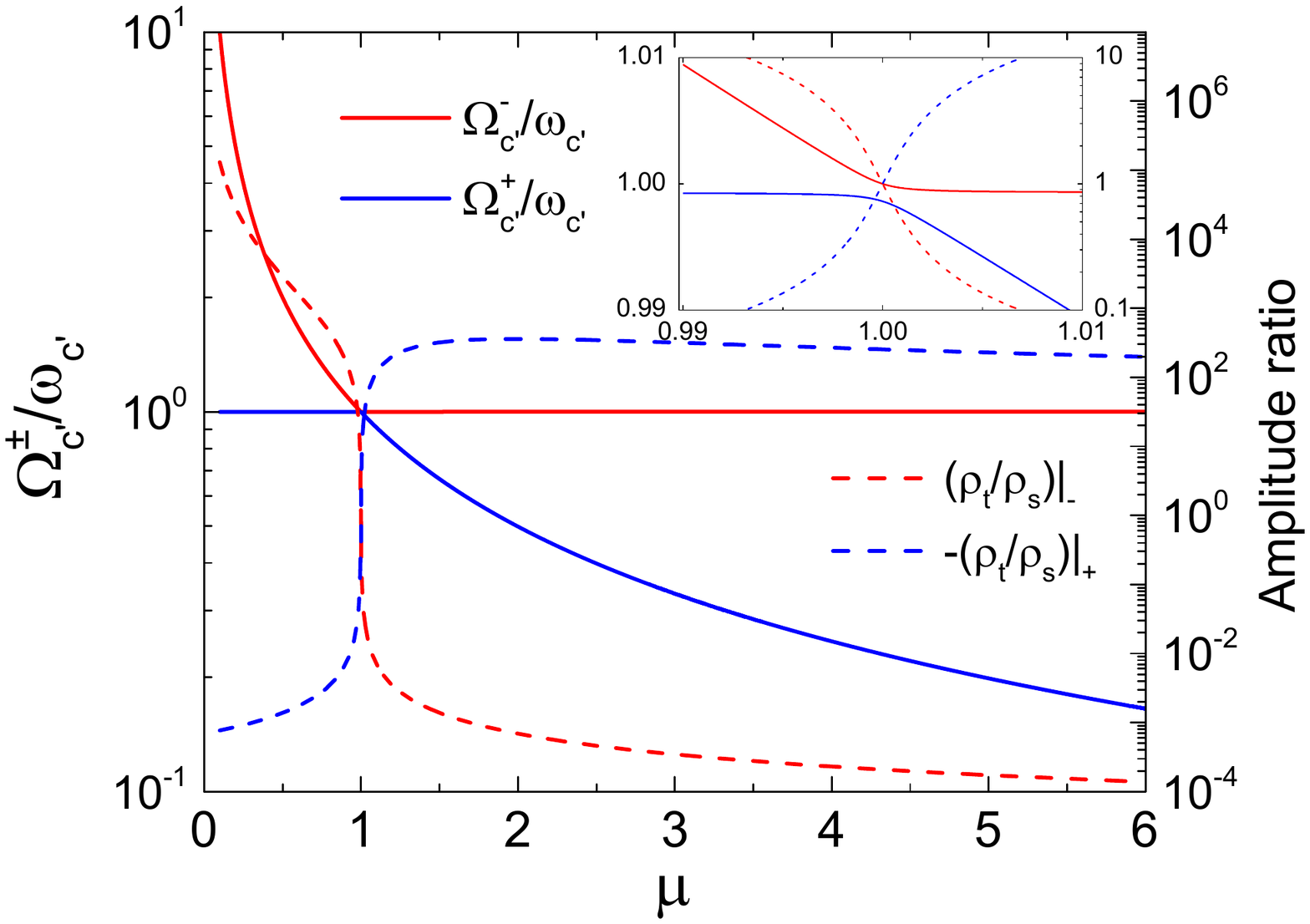}
\vspace{-6.5cm}
\caption{Eigenvalues and eigenvectors of the reduced-cyclotron motion of the two-ion crystal. The solid lines correspond to the frequencies in units of the reduced-cyclotron frequency of the sensor ion $\omega _{c^{\prime}}$, whereas the dashed lines correspond to the amplitude ratios (target over sensor). The inset shows a zoomed-in version around $\mu=1$.}
\label{fig::cyclotron}
\end{figure}

\begin{figure}
\centering
\includegraphics[width=0.5\textwidth]{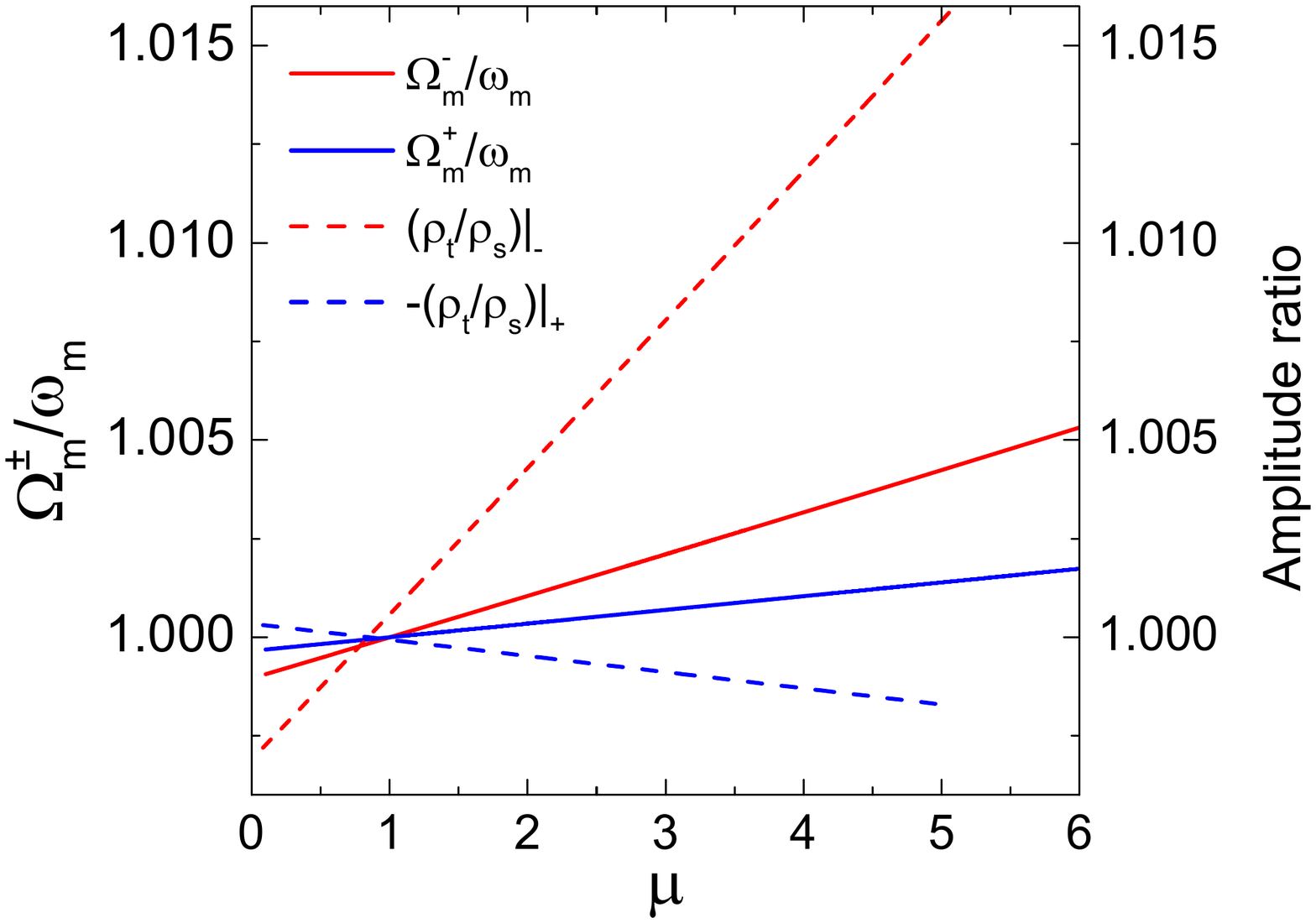}
\vspace{-6.5cm}
\caption{Eigenvalues and eigenvectors of the magnetron motion of the two-ion crystal. The solid lines show the frequencies normalized to their respective values at $\mu=1$.  $\omega _m$ equals 1860.84~kHz and 5591.06~kHz, for the common and the stretch mode, respectively (see Eq.~(\ref{eq::radial-freqs-homogeneous})). The dashed lines, on the other hand, show the amplitude ratios (target over sensor).}
\label{fig::magnetron}
\end{figure}

In the general case ($\mu \neq 1$) the motions are very similar; there are two in-phase (common) modes, and two out-of-phase (stretch) modes. The frequencies, however, are not the same; neither are the relative amplitudes of the two ions, which will not be equal anymore (see Fig.~\ref{fig::homogeneous-radial}(c)~and~(d)). In order to obtain these quantities, one uses the \emph{ansatz} $u_{t,s}=\rho_{t,s}e^{-i \Omega t}$ in Eq.~(\ref{eq::radial-matrix}). This yields for the radial eigenmotions

\begin{equation}
\left[
\Omega^2 
\left(
\begin{matrix}
1 & 0 \\
0 & 1
\end{matrix}
\right)
- \Omega \omega_{cs}
\left(
\begin{matrix}
\frac{1}{\mu} & 0 \\
      0       & 1
\end{matrix}
\right)
+
\omega_0^2
\left(
\begin{matrix}
\frac{1}{\mu} & -\frac{1}{2}\frac{1}{\mu} \\
 -\frac{1}{2} & 1
\end{matrix}
\right)
\right]
\left(
\begin{matrix}
\rho_t \\
\rho_s
\end{matrix}
\right)
= 0  \,,
\end{equation}

\noindent which is known as a non-linear eigenproblem. A numerical solution can be computed \cite{Octave1} to obtain the parameters of interest, that is, the motional frequencies $\Omega$ and amplitude ratios $\rho_t/\rho_s$ for each of the four eigenmotions. The results are presented in Figs.~\ref{fig::cyclotron}~and~\ref{fig::magnetron}.

\section{\label{sec::mass-meas}Mass measurements with two trapped ions and optical detection}
So far, calculations of the motional frequencies of a two-ion crystal as well as its dynamics have been shown. This allows determining the individual cyclotron frequencies for each of the ions forming the unbalanced crystal using a generalization of the invariance theorem (Eq.~(\ref{eq::invariance})) that reads \cite{jain2018,jain2019} 
\begin{equation}
\sum_{i=1}^6 \omega_i^2 = \omega_{cs}^2 + \omega_{ct}^2, \label{invariance_gen}
\end{equation}
\noindent where $i $ is an index to account for each of the eigenfrequencies. The measurement of the crystal eigenfrequencies require external fields applied in similar way as it is done in many experiments \cite{Corn1989,Koni1995,Ubie2009,Elis2013}. However, the detection will vary, i.e., by using optical photons from the laser-cooled $^{40}$Ca$^+$ ion, instead of a micro-channel plate detector or an electrical circuit. 

\subsection{Measurement scheme}

\begin{figure}[b]
\centering\includegraphics[width=1\linewidth]{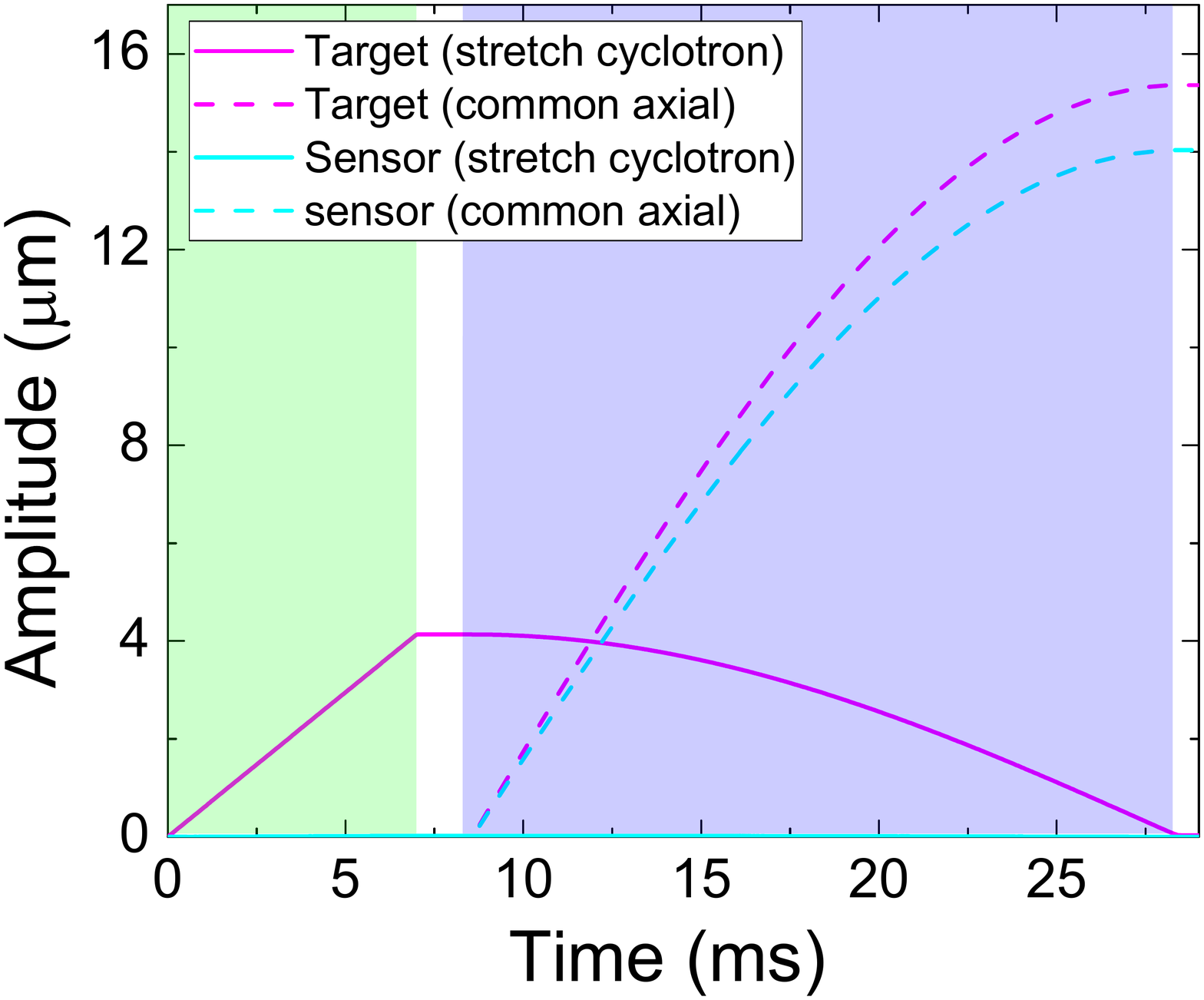}
\vspace{-0.7cm}
\caption{Evolution of the amplitude of the target and sensor ions' eigenmotions as a function of time. The target ion is $^{48}$Ca$^+$. First, a dipolar field in resonance with the stretch mode of the reduced-cyclotron motion ($\nu _{\hbox{\scriptsize{RF}}}=\Omega_{c^{\prime}}^+/2\pi$) is applied in the radial direction (green-shaded area). Subsequently, a quadrupolar field with frequency $\nu _{\hbox{\scriptsize{RF}}}=(\Omega_{c^{\prime}}^+ - \Omega_{z}^-)/2\pi$ is applied in the axial direction (blue-shaded area).  \label{Figure_conversion_2}}
\end{figure}

The axial motion of the crystal can be probed and monitored by applying simultaneously a resonant dipolar field and the cooling lasers. See for example Ref.~\cite{Drew2004} for two $^{40}$Ca$^+$ ions stored in a linear Paul trap. Moreover, a model to describe the evolution of the fluorescence profile as a function of the excitation frequency, has been developed \cite{Domi2017b}, yielding for the amplitude of the sensor ion in an unbalanced crystal
\begin{equation}
z_{s}=\frac{F_e}{(m_s+m_t)} \{ (2 \gamma_z \omega_{\rm dip})^2 + ((\Omega_z^-)^2 - \omega_{\rm dip}^2)^2 \}^{-1/2}\times \frac {z_{t}}{Z_-}, \label{fit_osc}
\end{equation}

\noindent where $F_e$ and $\omega_{\rm dip}$ are, respectively, the force and the frequency of the harmonic driving force, and $\gamma_z$ the damping coefficient. The values of these coefficients have to be determined experimentally as it was done for a $^{40}$Ca$^+$-$^{40}$Ca$^+$ crystal in an open-ring Paul trap \cite{Domi2017b}. Other eigenfrequency-detection schemes developed in single-ion experiments might be also utilized (see e.g. \cite{Knun2010,Knun2012}).

\begin{figure}[t]
\centering\includegraphics[width=1\linewidth]{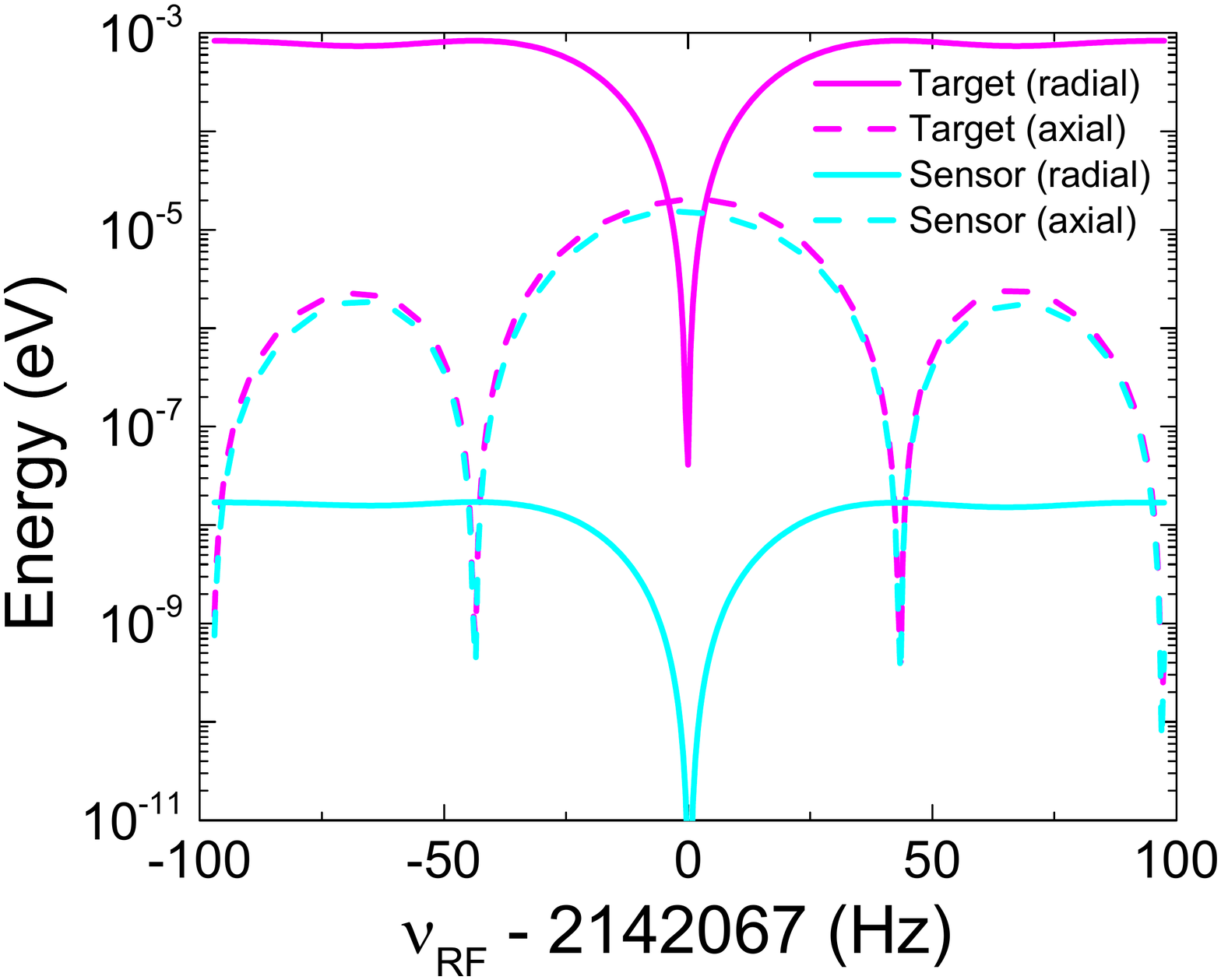}
\vspace{-0.7cm}
\caption{Energy of the target and sensor ions' eigenmotions when a $\pi$-pulse is applied around $\Omega_{c^{\prime}}^+ - \Omega_{z}^-$ for 20~ms (blue-shaded area in Fig.~\ref{Figure_conversion_2}). $\Omega_{c^{\prime}}^+ - \Omega_{z}^-= 2\pi \times 2142067$~Hz. The initial conditions for the target and sensor ions before conversion are also shown in Fig.~\ref{Figure_conversion_2} (right border of the green-shaded area). \label{Figure_conversion}}
\end{figure}

For the radial eigenfrequencies the measurement scheme is decided on after observing Figs.~\ref{fig::cyclotron}~and~\ref{fig::magnetron}. One can see in Fig.~\ref{fig::cyclotron} that for the common mode, when $\mu > 1$, the radius of the target ion is several orders of magnitude smaller than the radius of the sensor ion. This means that a small excitation of the crystal can be optically observed through the sensor ion if $m_t>m_s$. On the other hand, for the stretch mode, the situation is reversed, thus, it will be difficult to observe optically the excitation of the crystal through the $^{40}$Ca$^+$ ion. This will require a different scheme comprising the excitation, followed by the application of a $\pi$-pulse at $\Omega_{c^{\prime}}^+ - \Omega_{z}^-$ to convert reduced-cyclotron stretch to common axial \cite{Corn1990b}. Figure~\ref{Figure_conversion_2} shows the variation in amplitude of the target and sensor ions when a dipolar field is applied at $\Omega_{c^{\prime}}^+$ in the radial direction (green-shaded area). Only the amplitude of the target ion increases. The application of the $\pi$-pulse at $\Omega_{c^{\prime}}^+ - \Omega_{z}^-$ thereafter, results in a full conversion from the stretch reduced-cyclotron mode of the crystal to the common axial mode, where the oscillation amplitudes of the target and sensor ions are large, and thus the sensor ion can be monitored though its fluorescence photons. Figure~\ref{Figure_conversion} shows the energy of the target and sensor ions as a function of the frequency of the quadrupolar field applied in the axial direction. The radiofrequency field is applied during 20~ms. In this scenario the lasers are off during the time the radiofrequency fields are applied, and are interacting with the sensor ion afterwards for optical detection, following the scheme described in Ref.~\cite{Domi2017}. 

Regarding the magnetron modes, it is not possible to observe the magnetron stretch-mode ($\Omega _{m}^+$ ) directly, since the center of charge does not move and thus, it cannot be coupled to the dipolar field applied in the radial direction. Nevertheless, using Eqs.~(\ref{aux})~and~(\ref{eq::radial-freqs-homogeneous}), $\Omega _{m}^+=3.00458\times \Omega _{m}^-$ for a balanced crystal. In addition, from Fig.~\ref{fig::magnetron}, one can deduce that the differences between the magnetron eigenfrequencies of a balanced crystal and those of an unbalanced one, are negligible so that $\Omega _{m}^+$ does not need to be measured directly.

\subsection{Frequency shifts for large oscillation amplitudes}
\begin{figure}[t]
\centering\includegraphics[width=1\linewidth]{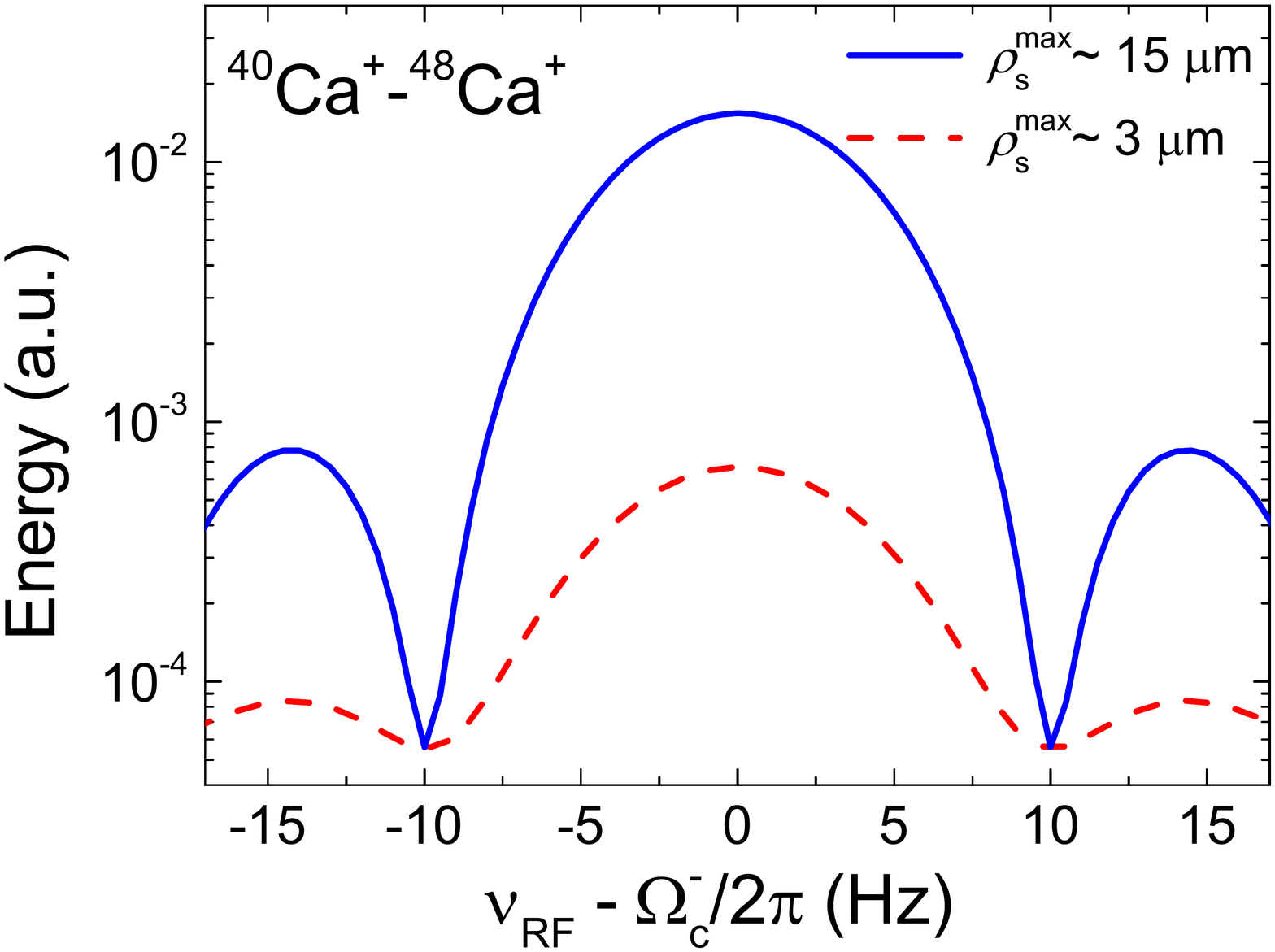}
\vspace{-0.85cm}
\caption{Evolution of the reduced-cyclotron common-mode energy of an unbalanced crystal as a function of the frequency of an external dipolar field around $\Omega _{c^{\prime}}^-$ applied in the radial direction. The low-amplitude approximation was not used. Most of the energy goes into the sensor ion. The excitation time was 100~ms. $\Omega _{c^{\prime}}^-/2\pi=2685282$~Hz. \label{sys_cyc}}
\end{figure}
Up to this stage, our calculations were based on the low-amplitude approximation (Eq.~(\ref{eq::linearized})). However, when the amplitude of one of the ions in the crystal becomes large, some of the frequencies of the crystal modes might shift with respect to the values given in Figs.~\ref{fig::axial},~\ref{fig::cyclotron}~and~\ref{fig::magnetron}. In order to determine these frequencies accurately, simulations have been carried out consisting in the application of time-varying dipolar electric fields. The frequencies are varied around the values obtained from the low-amplitude approximation, and the total energy gained by the crystal is observed. This procedure is similar to that of the final experiment, thus giving an accurate estimation of an eventual frequency shift due to the non-harmonicity of the Coulomb interaction. The field strength was varied to obtain different values of the final oscillation amplitudes of the ions.  

Figure~\ref{sys_cyc}  shows the energy gained in the reduced-cyclotron common-mode as a function of the radiofrequency of the external field. $\Omega _{c^{\prime}}^-$ remains unchanged for $\rho _s^{\hbox{\scriptsize{max}}}$ values of up to $15$~$\mu$m  ($\rho_t^{\hbox{\scriptsize{max}}}\simeq 50$~nm). For the axial common mode, $\Omega _{z}^-/2\pi=95152.61(5)$~Hz remains unchanged when $z_s^{\hbox{\scriptsize{max}}}$ and $z_t^{\hbox{\scriptsize{max}}}$ are $3.6$~$\mu$m and $3.9$~$\mu$m, respectively. There is a frequency shift $\Delta \Omega _{z}^-/2\pi =0.06(6)$~Hz when $z_s^{\hbox{\scriptsize{max}}}=7.2$~$\mu$m ($z_t^{\hbox{\scriptsize{max}}}=7.8$~$\mu$m), which increases further to 0.43(6)~Hz when $z_s^{\hbox{\scriptsize{max}}}=18$~$\mu$m and $z_t^{\hbox{\scriptsize{max}}}=19.6$~$\mu$m. No frequency shift has been observed for the magnetron common mode. 

The frequency of the reduced-cyclotron stretch mode $\Omega _{c^{\prime}}^+$ is neither shifted for the largest amplitude considered in this work; $\rho _t^{\hbox{\scriptsize{max}}}= 7.4$~$\mu$m ($\rho _s^{\hbox{\scriptsize{max}}}= 40$~nm). As written above, due to the very low amplitude of the sensor ion, this motion has to be observed in the axial direction after conversion (Fig.~\ref{Figure_conversion}). In such scheme, an amplitude of $\rho _t^{\hbox{\scriptsize{max}}}= 5$~$\mu$m is already sufficient  as can be inferred from Fig.~\ref{Figure_conversion_2}. The coupling to a mode with lower frequency has an amplitude-magnifying effect, which can be used to lessen the systematic effects that are amplitude dependent, for example the one arising from the non-harmonicity of the Coulomb interaction.

\begin{figure}[t]
\centering\includegraphics[width=1\linewidth]{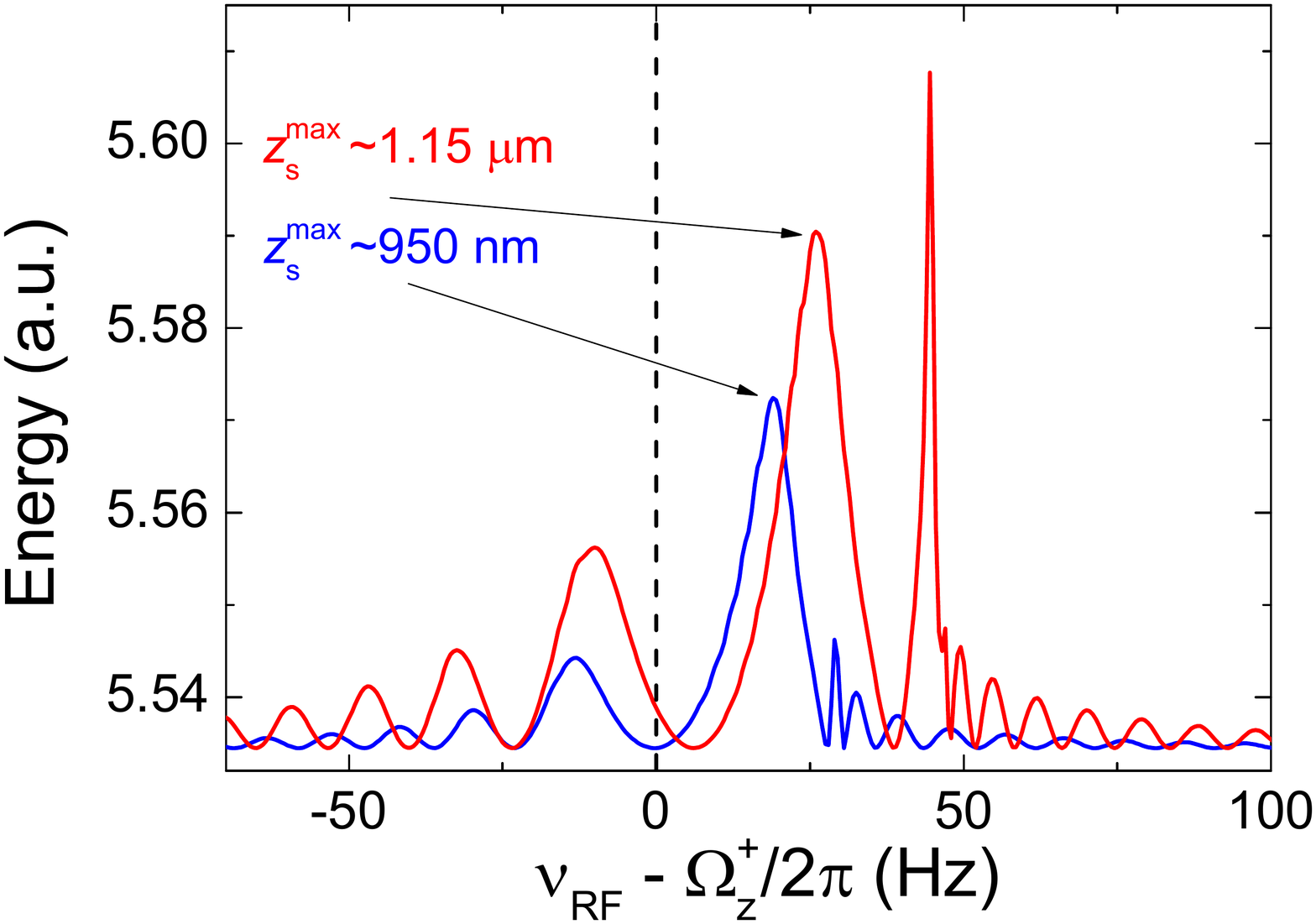}
\vspace{-0.95cm}
\caption{Evolution of the axial stretch-mode energy of the unbalanced crystal $^{40}$Ca$^+$-$^{48}$Ca$^+$ as a function of the frequency of an external dipolar field around $\Omega _{z}^+$ applied in the axial direction. The low-amplitude approximation was not used. The excitation time was 100~ms. $\Omega _{z}^+/2\pi=166173$~Hz. \label{sys_axial}}
\end{figure}

The largest effect is observed in $\Omega _z ^+$, since the ions move in opposite phase along the direction where their Coulomb interaction is stronger. The effect is shown in Fig.~\ref{sys_axial}, where a significant shift and an important deviation from the expected profile can be observed for very low oscillation amplitudes; $\Delta \Omega _z ^+/2\pi\simeq19$~Hz when $z _s^{\hbox{\scriptsize{max}}}= 950$~nm, and $\Delta \Omega _z ^+/2\pi\simeq26$~Hz when $z_s^{\hbox{\scriptsize{max}}}= 1.16$~$\mu$m. This means that the precise measurement of the axial stretch mode will require reaching the regime of low vibrational quantum number near the ground state, i.e., by performing side-band cooling \cite{Good2016}, or alternatively one could read the motional changes through another eigenmode, for example making a conversion from stretch axial to common magnetron, again exploiting the amplitude-magnifying effect.

\section{\label{sec::conclusions}Conclusions and perspectives}
In this paper we have shown results from calculations of the dynamics of an unbalanced two-ion crystal for applications in high-performance mass spectrometry. The main issue of having two ions in the same trap potential, which might perturb mass measurements, is the Coulomb interaction. This has been studied in detail to obtain the eigenfrequencies and eigenvectors of the crystal. The frequencies can be measured by applying external radiofrequency fields in resonance with each of the modes and observing the sensor ion motion through the fluorescence photons. $\Omega _{c^{\prime}}^-$, $\Omega _{z}^-$ and $\Omega _{m}^-$ can be measured directly using axial and radial laser-cooling beams. $\Omega _{m}^+$ is determined with sufficient precision through $\Omega _{m}^-$. $\Omega _{c^{\prime}}^+$, has to be measured through $\Omega _{z}^-$, and $\Omega _{z}^+$ measured through $\Omega _{m}^-$, the latter provided one cannot work in the low vibrational regime close to the ground state. The novelty of our method lies also in the optical detection \cite{Rodr2012}, allowing monitoring the sensor ion permanently in the trap, even when probing some of the crystal's eigenmodes. Still, one has to experimentally verify how accurately one can determine the frequencies from the fluorescence measurements in a Penning trap, as this issue has only been investigated in Paul traps \cite{Drew2004,Domi2017}.

\begin{figure}[t]
\centering\includegraphics[width=1\linewidth]{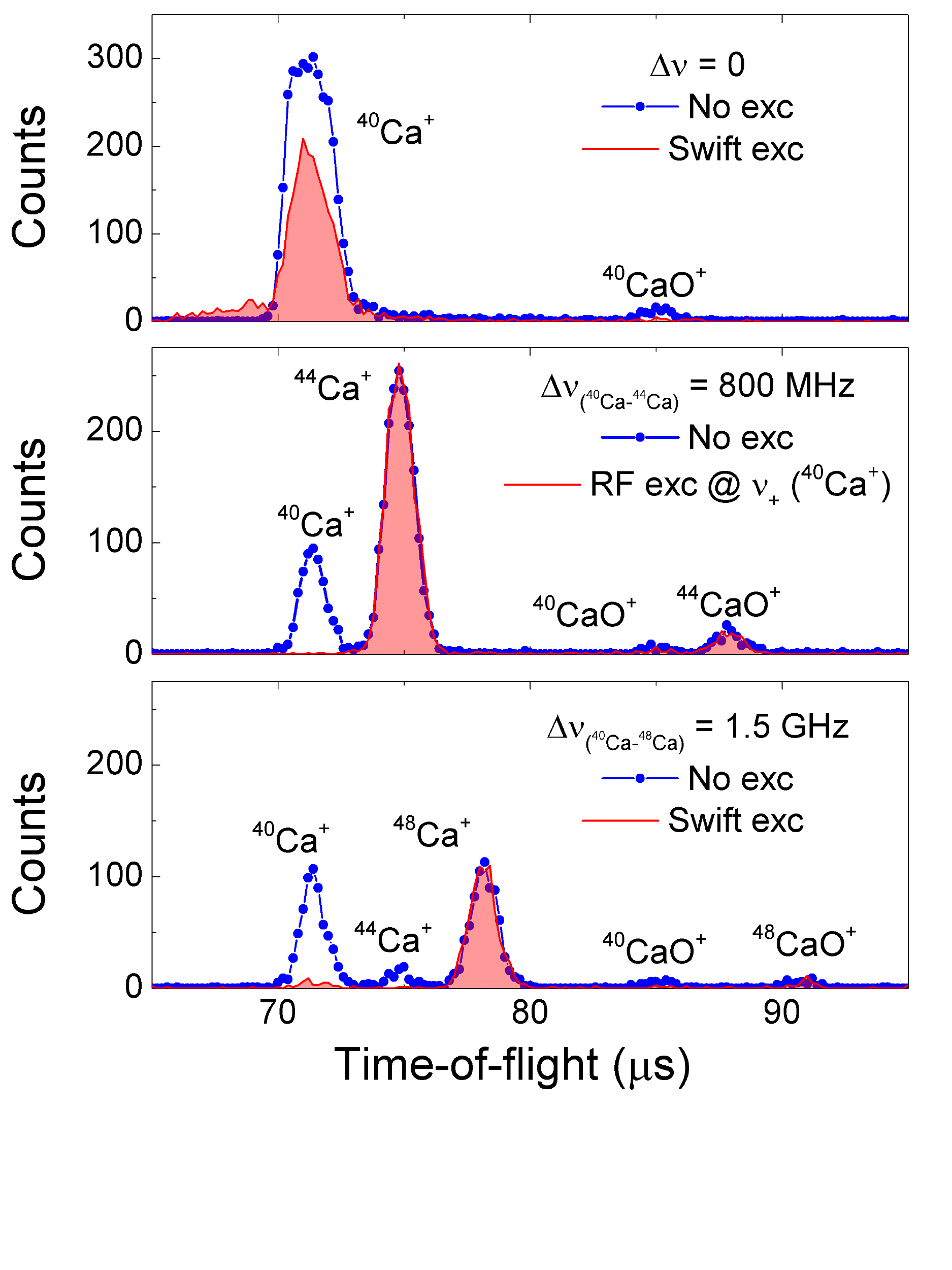}
\vspace{-2.2cm}
\caption{Time-of-flight spectra for calcium isotopes produced by photoionization in the open-ring Penning trap. The setup is described in Ref.~\cite{Guti2019}. $\Delta \nu $ is the frequency difference with respect to $\nu _{\hbox{S}_0 \rightarrow \hbox{P}_1}$($^{40}$Ca$^+$), due to the isotopic shift. Further details are given in the text. \label{calcium_tof}}
\end{figure}

A first experiment not only to prove this concept but also to quantify systematic effects, can be performed with the TRAPSENSOR facility \cite{Guti2019} using unbalanced crystals comprising two calcium isotopes. The simultaneous production of different calcium isotopes inside the trap has been already demonstrated. Figure~\ref{calcium_tof} shows three time-of-flight spectra containing each of them, $^{40}$Ca$^+$ ions and ions from other calcium isotopes. The ions are generated by photoionization using a laser beam, the frequency of which is shifted to drive the $^1$S$_0 \rightarrow ^1$P$_1$ transition ($\lambda =422$~nm) of the isotope of interest. This beam is superimposed to another one with $\lambda =375$~nm. In order to remove unwanted species, and select the target ion, a dipolar field at fixed frequency or a Stored-waveform-inverse Fourier-transform (SWIFT) pulse was applied \cite{Guan1996}. The lower panel of Fig.~\ref{calcium_tof} demonstrates the production of $^{40}$Ca$^+$ and $^{48}$Ca$^+$  in the same amount, without ions from any of the other naturally abundant calcium isotopes. These measurements with the calcium isotopes will serve also to validate Eq.~(\ref{invariance_gen}), since the measurement of the free cyclotron frequencies for $^{42}$Ca$^+$, $^{44}$Ca$^+$ and $^{48}$Ca$^+$ can be carried out in the same way as for $^{40}$Ca$^+$, just by properly detuning the wavelengths of the cooling lasers to account for the isotopic shifts.

This new measurement technique is of particular interest for the investigation of rare exotic particles that can only be provided in minute quantities, often only single atoms at a time. This concerns for example superheavy nuclides \cite{Bloc2019}. In this case an efficient capture of the ion, that has been created outside the trap, is necessary. Such experiments require also sympathetic cooling of the SHE with the laser-cooled $^{40}$Ca$^+$ ion. At present, heavy ions have been already captured in the open-ring trap \cite{Guti2019}, and the cooling of a few $^{40}$Ca$^+$ ions produced outside this trap, has been demonstrated. Additional applications of this new approach may arise once this cooling has been accomplished. For example, the formation of cold molecular ions of exotic superheavy atoms would be of interest.
\section*{Acknowledgement}
M.J.G., J.B., F.D., M.B. and D.R. acknowledge support from the Spanish MINECO (now MCIU) through the project  FPA2015-67694-P, from the Spanish Ministry of Education through PhD fellowships FPU15-04679 and FPU17/02596, and from the University of Granada "Plan propio - Programa de Intensificaci\'on de la Investigaci\'on", project PP2017-PRI.I-04. I.A. and E.S. acknowledge support from Grant No. PGC2018-095113-B-I00 (MCIU/AEI/FEDER,UE), from Basque Government though project IT986-16 and Ph.D. Grant No.PRE-2015-1-039 (I.A.) and from QMiCS (820505) and OpenSuperQ (820363) of the EU Flagship on Quantum Technologies, and EU FET Open Grant Quromorphic. The construction of the facility (experimental results shown in Fig.~\ref{calcium_tof}), was supported by the European Research Council (contract number 278648-TRAPSENSOR), the above-mentioned project FPA2015-67694-P, project FPA2012-32076, and the infrastructure projects UNGR10-1E-501 and UNGR13-1E-1830 (MINECO/FEDER/Universidad de Granada), and IE-5713 (Junta de Andalucía/FEDER). 
\appendix
\section{Equations of motion} \label{Equations_of_motion}
The equations of motion of two singly charged ions of masses $m_t$ and $m_s$ confined simultaneously in a Penning trap, using coordinates with respect to the trap center, can be written as

\begin{eqnarray}
\begin{cases}
\displaystyle{\ddot{x}_t = \frac{1}{\mu} \frac{ \omega_0^2 }{2} x_t + \frac{1}{\mu} \omega_{cs} \dot{y}_t +
\frac{1}{m_t} \frac{e^2}{4 \pi \varepsilon_0} \frac{ \left( \vec r_t - \vec r_s \right) \cdot \hat x }{ \left| \vec r_t - \vec r_s \right|^3 }}
\\
\displaystyle{\ddot{y}_t = \frac{1}{\mu} \frac{ \omega_0^2 }{2} y_t - \frac{1}{\mu} \omega_{cs} \dot{x}_t +
\frac{1}{m_t} \frac{e^2}{4 \pi \varepsilon_0} \frac{ \left( \vec r_t - \vec r_s \right) \cdot \hat y }{ \left| \vec r_t - \vec r_s \right|^3 }}
\\
\displaystyle{\ddot{z}_t = -\frac{1}{\mu} \omega_0^2 z_t +
\frac{1}{m_t} \frac{e^2}{4 \pi \varepsilon_0} \frac{ \left( \vec r_t - \vec r_s \right) \cdot \hat z }{ \left| \vec r_t - \vec r_s \right|^3 }}
\\
\displaystyle{\ddot{x}_s = \frac{ \omega_0^2 }{2} x_s + \omega_{cs} \dot{y}_s -
\frac{1}{m_s} \frac{e^2}{4 \pi \varepsilon_0} \frac{ \left( \vec r_t - \vec r_s \right) \cdot \hat x }{ \left| \vec r_t - \vec r_s \right|^3 }}
\\
\displaystyle{\ddot{y}_s = \frac{ \omega_0^2 }{2} y_s - \omega_{cs} \dot{x}_s -
\frac{1}{m_s} \frac{e^2}{4 \pi \varepsilon_0} \frac{ \left( \vec r_t - \vec r_s \right) \cdot \hat y }{ \left| \vec r_t - \vec r_s \right|^3 }}
\\
\displaystyle{\ddot{z}_s = - \omega_0^2 z_s -
\frac{1}{m_s} \frac{e^2}{4 \pi \varepsilon_0} \frac{ \left( \vec r_t - \vec r_s \right) \cdot \hat z }{ \left| \vec r_t - \vec r_s \right|^3 }}
\end{cases} \,.
\label{eq::full-old-coords}
\end{eqnarray}

\noindent where $\mu=m_t/m_s$. The parameters $\omega_{zt} = \omega_0/\sqrt{\mu} $ and $ \omega_{ct} = \omega_{cs} / \mu $, are the axial and true cyclotron frequencies of the target ion, respectively. $\omega_0$ and $\omega_{cs}$ are the axial and true cyclotron frequency of the sensor ion. The last term of each equation accounts for the Coulomb repulsion between the two ions.

In the case of an unbalanced crystal oriented in the axial direction (see Fig.~\ref{fig::axial1}), the equilibrium positions are $ z_t^{eq} = -z_s^{eq} = d/2 $, where $d^3 = e^2/\left( 2\pi \varepsilon_0 m_s \omega_0^2 \right)$ is the equilibrium distance. Referring the coordinates of each ion to its equilibrium position, the resulting equations are

\begin{eqnarray}
\begin{cases}
\displaystyle{\ddot{x}_t = \frac{1}{\mu} \frac{ \omega_0^2 }{2} x_t + \frac{1}{\mu} \omega_{cs} \dot{y}_t +
 \frac{ \omega_0^2 d^3 }{2} \frac{1}{\mu} \frac{ \left( \vec r_t - \vec r_s \right) \cdot \hat x }{ \left| \vec r_t - \vec r_s + d \hat z \right|^3 }}
\\
\displaystyle{\ddot{y}_t = \frac{1}{\mu} \frac{ \omega_0^2 }{2} y_t - \frac{1}{\mu} \omega_{cs} \dot{x}_t +
 \frac{ \omega_0^2 d^3 }{2} \frac{1}{\mu} \frac{ \left( \vec r_t - \vec r_s \right) \cdot \hat y }{ \left| \vec r_t - \vec r_s + d \hat z \right|^3 }}
\\
\displaystyle{\ddot{z}_t = -\frac{1}{\mu} \omega_0^2 \left( z_t + \frac{d}{2} \right) +
 \frac{ \omega_0^2 d^3 }{2} \frac{1}{\mu} \frac{ \left( \vec r_t - \vec r_s \right) \cdot \hat z }{ \left| \vec r_t - \vec r_s + d \hat z \right|^3 }}
\\
\displaystyle{\ddot{x}_s = \frac{ \omega_0^2 }{2} x_s + \omega_{cs} \dot{y}_s -
 \frac{ \omega_0^2 d^3 }{2} \frac{ \left( \vec r_t - \vec r_s \right) \cdot \hat x }{ \left| \vec r_t - \vec r_s + d \hat z \right|^3 }}
\\
\displaystyle{\ddot{y}_s = \frac{ \omega_0^2 }{2} y_s - \omega_{cs} \dot{x}_s -
 \frac{ \omega_0^2 d^3 }{2} \frac{ \left( \vec r_t - \vec r_s \right) \cdot \hat y }{ \left| \vec r_t - \vec r_s + d \hat z \right|^3 }}
\\
\displaystyle{\ddot{z}_s = - \omega_0^2 \left( z_s - \frac{d}{2} \right) -
 \frac{ \omega_0^2 d^3 }{2} \frac{ \left( \vec r_t - \vec r_s \right) \cdot \hat z }{ \left| \vec r_t - \vec r_s + d \hat z \right|^3 }}
\end{cases}
\label{eq::full}
\end{eqnarray}

This system of equations can be partially solved for $\mu=1$ (equal masses) for the center of mass motion, since the Coulomb term cancels out for $\Xi _{\hbox{\scriptsize{COM}}}= \xi_t - \xi_s, \,\,\xi=\left\{ x,y,x \right\} $. The resulting motion is the same as for the single ion. 
\newpage
 %

\end{document}